\documentclass[showpacs,aps,amssymb,pra,amsmath]{revtex4}
\usepackage{float}
\usepackage{epsfig}
\usepackage{bm}
\usepackage{amsmath,graphicx}
\usepackage{subfig}
\usepackage{setspace}
\usepackage{color}

\newcommand{\beq}{\begin{equation}}
\newcommand{\eeq}{\end{equation}}
\newcommand{\bea}{\begin{eqnarray}}
\newcommand{\eea}{\end{eqnarray}}
\def\tx{{\tilde{x}}}
\def\tq{{\tilde{q}}}
\def\ty{{\tilde{y}}}
\def\talpha{{\tilde{\alpha}}}
\def\bk{{\bf k}}
\def\ba{{\bf a}}

\def\bq{{\bf q}}

\def\br{{\bf r}}

\begin{document}
\title{Density-functional theory for the crystalline phases of a two-dimensional dipolar Fermi gas}

\author{B. P. van Zyl, W. Kirkby, and W. Ferguson}
\affiliation{Department of Physics, St. Francis Xavier University, Antigonish, Nova Scotia, Canada
B2G 2W5}

\date{\today}
\begin{abstract}
Density-functional theory is utilized to investigate the zero-temperature 
transition from a Fermi liquid to an inhomogeneous stripe, or Wigner crystal phase, predicted to occur in a one-component,
spin-polarized,  two-dimensional dipolar Fermi gas.   Correlations are treated semi-exactly within the local-density approximation  using
an empirical fit to Quantum Monte Carlo data.
We find that the inclusion of the nonlocal contribution to the Hartree-Fock energy is crucial for the onset of an instability to an inhomogeneous
density distribution.  Our density-functional theory supports a transition to both a one-dimensional stripe phase, and
a triangular Wigner crystal.  However, we find that there is an instability first to the stripe phase, followed by a transition to the Wigner crystal
at higher coupling.
\end{abstract}
\pacs{67.85.Lm,~64.70.D,~71.45.Gm,~31.15.xt}
\maketitle

\section{Introduction}\label{intro}
In this paper, we consider a strictly two-dimensional (2D), spin-polarized, Fermi gas interacting {\it via} an
isotropic,  repulsive dipolar interaction ({\it i.e.,} all of the moments are aligned parallel to the $z$-axis), {\it viz.,}
\beq\label{Vdd}
V_{\rm dd}(\br - \br') = \frac{C_{\rm dd}}{4\pi |\br - \br'|^3}~,
\eeq
where $C_{\rm dd}=\mu_0 d^2$, $d$ is the magnetic dipole moment of an atom (which we take to be charge neutral,
{\it e.g.,} $^{161}$~Dy with $d \sim 10 \mu_B$), and $\br$ and $\br'$ are
coordinates in the 2D $x-y$ plane.
It will prove useful later to define the following additional quantities:
$r_0 = M C_{\rm dd}/(4 \pi \hbar^2)$, $k_{\rm F} = \sqrt{4 \pi \rho}$, and
$\lambda = k_{\rm F} r_0$.  Here, $M$ is the mass of an atom, $\rho$ is the 2D density, and $k_{\rm F}$ is  the 2D Fermi wave vector.

The above system has received considerable theoretical attention over the last few years (see, {\it e.g}., Refs.~\cite{Lin10,Cremon10,Sun10,Yamaguchi10,Sieberer11,Block12,Babadi11,block2014,Parish12,Mateeva12,abed12,babadi2013}),
 owing to the possibility of experimentally observing
the quantum phase transition from the normal Fermi liquid (FL) to an ordered state, {\it e.g.,} a one-dimensional
stripe phase (1DSP) or triangular Wigner crystal (WC).  
The basic idea is that because of the ``long range'' $r^{-3}$ repulsive potential in 2D, for a sufficiently large value of the dipole moment (equivalently, density),
it will be energetically favourable for the system to spontaneously break  translational invariance to an inhomogeneous phase.
In fact, to date, there is an unresolved controversy in the literature
when it comes to answering the question of which inhomogeneous phase the 2D dipolar Fermi gas (dFG) 
spontaneously possesses above some
critical coupling.  

Early calculations within the random
phase approximation suggested a transition to a 1DSP at $k_{\rm F} r_0 \approx 0.61$~\cite{Sun10,Yamaguchi10}, while improvements including the so-called
STLS scheme yield $k_{\rm F}r_0 \approx 6$~\cite{Parish12}.  More sophisticated investigations employing 
the conserving Hartree-Fock (HF) approximation point to a 1DSP transition occurring at $k_{\rm F}r_0 \approx 1.4$~\cite{Sieberer11,Block12,Babadi11,block2014}.
Finally, utilizing variational
and Quantum Monte Carlo (QMC) techniques, a transition to a triangular WC phase at 
$k_{\rm F}r_0= 29\pm 4$~\cite{babadi2013} and $k_{\rm F}r_0 = 25 \pm 3$~\cite{Mateeva12,abed12}, respectively, is predicted to precede the formation of a
1DSP.  
Given the differing predictions of the nature of the ordered phase, and the critical density at which the system undergoes the
transition, we feel that there is ample motivation to present yet another theoretical approach to the problem; in the present work, our method of choice is
the density-functional theory (DFT)~\cite{DFT}. 

Density-functional theory has already been successfully applied to study  the Wigner crystalline phase in the degenerate 2D~\cite{vignale95,ghosh95,seidl99}, and
3D electron gas~\cite{Shore78,das88}, and has
recently been implemented to study  the equilibrium and collective excitations of a harmonically trapped, 2D dFG~\cite{fang,vanZyl12,vanZyl14}, as well as the study
of (classical) crystallization of magnetic dipolar monolayers in two-dimensions~\cite{teeffelen08} .  
It is therefore quite reasonable to believe that an application of DFT to study the quantum phase transition discussed above will likewise be fruitful.  
Surprisingly, to our knowledge, no such investigation specifically dealing with a degenerate 2D dFG  has been performed in the literature.  We propose to fill this gap by presenting a
DFT which will allow us to weigh in on the nature of the transition,
the critical interaction strength at which the transition occurs, as well as providing a useful test of the various density functionals recently developed in the context of the 2D dFG~\cite{abed12,fang,vanZyl12,vanZyl14}.

The rest of our paper is organized as follows.  In the next section, we develop a DFT for the study of the $T=0$ 2D dFG.  Section~\ref{results} presents our results for the
onset, and nature of the liquid-to-ordered phase.  Finally, in Sec.~\ref{closing}, we present our conclusions and closing
remarks.


\section{Density functional theory of the 2D dipolar Fermi gas}\label{2dDFT}
At the heart of DFT is the construction of the total energy of the system, which is a unique functional
of the one-body density, $\rho(\br)$, {\it viz.,}
\beq\label{Etot1}
E[\rho] = K[\rho] + E_{\rm int}[\rho] + E_{\rm ext}[\rho]~. 
\eeq
In Eq.~\eqref{Etot1}, $K[\rho]$ is the non-interacting kinetic energy (KE) functional of the system,
$E_{\rm int}[\rho]$ incorporates all of the quantum many-body interactions,  and
\beq\label{Eext}
E_{\rm ext}[\rho] = \int d^2r~\rho(\br) v_{\rm ext}(\br)~,
\eeq
is the energy functional associated with the external potential, $v_{\rm ext}(\br)$,
imposed on the system.  A variational minimization of Eq.~\eqref{Etot1} with respect to the density, $\rho(\br)$, leads to a description of the
zero-temperature ($T=0$) ground state properties of the  many-body system.   For an arbitrary inhomogeneous Fermi gas, the first two functionals in
Eq.~\eqref{Etot1} are not generally known.
However, for a uniform, {\it i.e.,} $v_{\rm ext}(\br)=0$, spin-polarized  2D Fermi gas at $T=0$, the non-interacting KE is known exactly, {\it viz.,}
\beq\label{2dke}
K[\rho_0] = \pi \frac{\hbar^2}{M} \int d^2r \rho_0^2~,
\eeq
where $\rho_0$ is the uniform density.
If the system is weakly inhomogeneous, it is reasonable to assume that  Eq.~\eqref{2dke} is still 
approximately valid, but with
$\rho_0 \to \rho(\br)$; this is the so-called local-density approximation (LDA), in which Eq.~\eqref{2dke} is known as the Thomas-Fermi (TF) KE functional.  

Using the definitions introduced in Sec.~\ref{intro}, Eq.~\eqref{2dke} reads
\beq\label{2dkelambda}
K[\lambda] = \frac{1}{16\pi} \frac{\hbar^2}{M r_0^4}\int d^2r~\lambda^4~.
\eeq
For a uniform system, $\lambda \to \lambda_0 = \sqrt{4\pi \rho_0}r_0$, while for the
inhomogeneous system, $\lambda \to \lambda(\br) = \sqrt{4 \pi \rho(\br)}r_0$.

The $T=0$ interaction energy functional, $E_{\rm int}[\rho]$, 
for a uniform, spin-polarized 2D dFG
is also known 
semi-exactly~\cite{fang,vanZyl12,abed12}.  In particular, one can decompose $E_{\rm int}[\rho]$ into
\beq\label{Eint}
E_{\rm int}[\rho] = E^{(1)}_{\rm dd}[\rho] + E_{\rm corr}[\rho]~,
\eeq
where the first term  in Eq.~\eqref{Eint} is the HF  energy, and 
the last term takes into account the quantum many-body correlations. The HF energy reads
\beq\label{E1dd}
E^{(1)}_{\rm dd}[\lambda] = \frac{8}{45 \pi^2}\frac{\hbar^2}{M r_0^4}\int d^2r~ \lambda^5~,
\eeq
while the correlation energy is obtained using an empirical fit to QMC
data presented in Ref.~\cite{abed12}, namely,
\beq\label{2dcorr}
E_{\rm corr}[\lambda] = -\frac{1}{32\pi}\frac{\hbar^2}{M r_0^4} \int d^2r~\lambda^6\ln \left( 1 + \frac{1}{a\sqrt{\lambda} + b \lambda + c \lambda^{\frac{3}{2}}}\right)~,
\eeq
where $a = 1.1958$, $b=1.1017$, and $c=-0.0100$. Equation~\eqref{2dcorr} 
may be viewed as being semi-exact up to $\lambda_0=70$.

Putting everything together, a DFT for an {\em inhomogeneous} 2D dFG may be constructed through a
standard
application of the LDA, $\lambda \to \lambda(\br)$, to the functionals of the uniform system, {\it viz.,}
\bea\label{EtotLDA}
E[\lambda(\br)] &=& \frac{1}{16\pi} \frac{\hbar^2}{M r_0^4}\int d^2r~\lambda(\br)^4 + 
\frac{8}{45 \pi^2}\frac{\hbar^2}{M r_0^4}\int d^2r~ \lambda(\br)^5\nonumber \\
& -&\frac{1}{32\pi}\frac{\hbar^2}{M r_0^4} \int d^2r~\lambda(\br)^6\ln \left( 1 + \frac{1}{a\sqrt{\lambda(\br)} + b \lambda(\br) + c \lambda(\br)^{\frac{3}{2}}}\right)~.
\eea
Note that for a uniform system, Eq.~\eqref{EtotLDA} is expected to be very accurate~\cite{abed12}. 

\section{Results}\label{results}

Not surprisingly, Eq.~\eqref{EtotLDA} has been suggested as a promising candidate for investigating  the 
quantum phase transition from the FL to an inhomogeneous ordered phase~\cite{abed12}.  To this end, we define the following quantity~\cite{ghosh95},
\beq\label{deltaE}
\Delta \varepsilon = \varepsilon_{\rm inhomo} - \varepsilon_{\rm uniform} = \frac{E[\lambda(\br)] - E[\lambda_0]}{\int d^2r~\rho_0}~,
\eeq
which represents the difference in energy (per particle) between the inhomogeneous and 
uniform phases.  We will adopt the notation that $\varepsilon$ always corresponds to an energy per particle, 
scaled by $\hbar^2/Mr_0^2$.
To proceed, 
we evaluate Eq.~\eqref{deltaE} by considering two different 
representations for the weakly inhomogeneous density distribution.  In our first case, we take the exceedingly simple form,
\beq\label{denmod}
\rho(\br) = \rho_0(1 + \alpha \cos(\bq\cdot \br))~,
\eeq
\beq\label{lambdamod}
\lambda(\br) = \lambda_0(1+\alpha\cos(\bq \cdot \br))^{\frac{1}{2}}~,
\eeq
which is suitable for studying, {\it e.g.,} a 1D modulated density profile.  We also consider 
a density modulation that mimics a 2D triangular lattice, {\it viz.,}
\beq\label{denmodWC}
\rho(\br)=\rho(x,y) =\rho_0 \left[\sqrt{1-\frac{3}{2}\alpha^2}+ \alpha \cos\left(q x\right)  + 2 \alpha \cos\left(\frac{q}{2}x\right) \cos\left(\frac{\sqrt{3}}{2}q y\right)   \right]^2~,
\eeq
\beq\label{lambdaWC}
\lambda(\br)=\lambda(x,y) =\lambda_0  \left[\sqrt{1-\frac{3}{2}\alpha^2}+ \alpha \cos\left(q x\right)  + 2 \alpha \cos\left(\frac{q}{2}x\right) \cos\left(\frac{\sqrt{3}}{2}q y\right)  \right]~.
\eeq
In the above, $\alpha \ll 1$ characterizes the amplitude of the density modulation around 
the uniform density $\rho_0$, with $\alpha=0$ corresponding to the liquid state.  Note that $\rho_0 = \int d^2r\rho(r)/\int d^2r$ in both Eqs.~\eqref{denmod} and \eqref{denmodWC}.

Owing to the fact that  
$\alpha \ll 1$, we may take a perturbative approach, and only consider Eq.~\eqref{deltaE} up to ${\cal O}(\alpha^2)$.  Using Eq.~\eqref{lambdamod} in the functionals defined in Sec.~\ref{2dDFT}, we obtain
\bea\label{deltaE2}
\frac{\Delta {\varepsilon}}{\alpha^2} &=& \frac{1}{8}\lambda_0^2 +
 \frac{2}{3\pi}\lambda_0^3
 -\frac{1}{8}\lambda_0^4 \left[ \frac{3}{2} \ln [f_0] + \frac{11}{16}\lambda_0 A + \frac{1}{16} \lambda_0^2 B\right]\nonumber \\
 &\equiv& \frac{1}{\alpha^2} ( \Delta \varepsilon_{\rm TF} + \Delta \varepsilon^{(1)}_{\rm dd} + \Delta \varepsilon_{\rm corr})~,
 \eea
where
\bea
f(\lambda) &=& \left( 1 + \frac{1}{a\sqrt{\lambda} + b \lambda + c \lambda^{\frac{3}{2}}}\right)~,
\\
f_0 &\equiv& f(\lambda_0),~~~
f'_0 \equiv \left. \frac{df}{d\lambda}\right |_{\lambda=\lambda_0},~~~
f''_0 \equiv \left. \frac{d^2f}{d\lambda^2}\right |_{\lambda=\lambda_0}~,\\
A&=& \frac{f_0'}{f_0}~,\\
B&=&\frac{f''_0f_0-f_0'^2}{f_0^2}~.
\eea
Similarly, inserting Eq.~\eqref{lambdaWC} into the functionals above, we get
\bea\label{deltaE2WC}
\frac{\Delta {\varepsilon}}{\alpha^2} &=& \frac{3}{2}\lambda_0^2 +
 \frac{8}{\pi}\lambda_0^3
 -\frac{1}{4}\lambda_0^4 \left[ 9 \ln [f_0] + \frac{33}{8}\lambda_0 A + \frac{3}{8} \lambda_0^2 B\right]\nonumber \\
 &\equiv& \frac{1}{\alpha^2} ( \Delta \varepsilon_{\rm TF} + \Delta \varepsilon^{(1)}_{\rm dd} + \Delta \varepsilon_{\rm corr})~.
 \eea
We remind the reader that all energies, $\varepsilon$, are per particle, and scaled by $\hbar^2/M r_0^2$ so that {\it e.g.,} $\Delta \varepsilon$,
is dimensionless.  
The terms on the right-hand side of Eqs.~\eqref{deltaE2} and \eqref{deltaE2WC} arise from Eq.~\eqref{2dkelambda},
Eq.~\eqref{E1dd}, and Eq.~\eqref{2dcorr}, respectively.
It is important to note that $\Delta \varepsilon$ is {\em independent of} $q$ in both modulated density scenarios, but this is a general result
for any small amplitude, periodic modulation to the liquid state. 
Consequently, it must be the case that neither Eqs.~\eqref{deltaE2} nor \eqref{deltaE2WC} allow for a transition ({\it i.e.,} $\Delta \varepsilon$ crossing through zero) to an 
inhomogeneous state, as that
would imply that the FL is unstable to an arbitrary density fluctuation.  
We can confirm this assertion numerically by examining Eq.~\eqref{deltaE2} (similar results follow from 
Eq.~\eqref{deltaE2WC}).
\begin{figure}[ht]
\centering \scalebox{0.4}
{\includegraphics{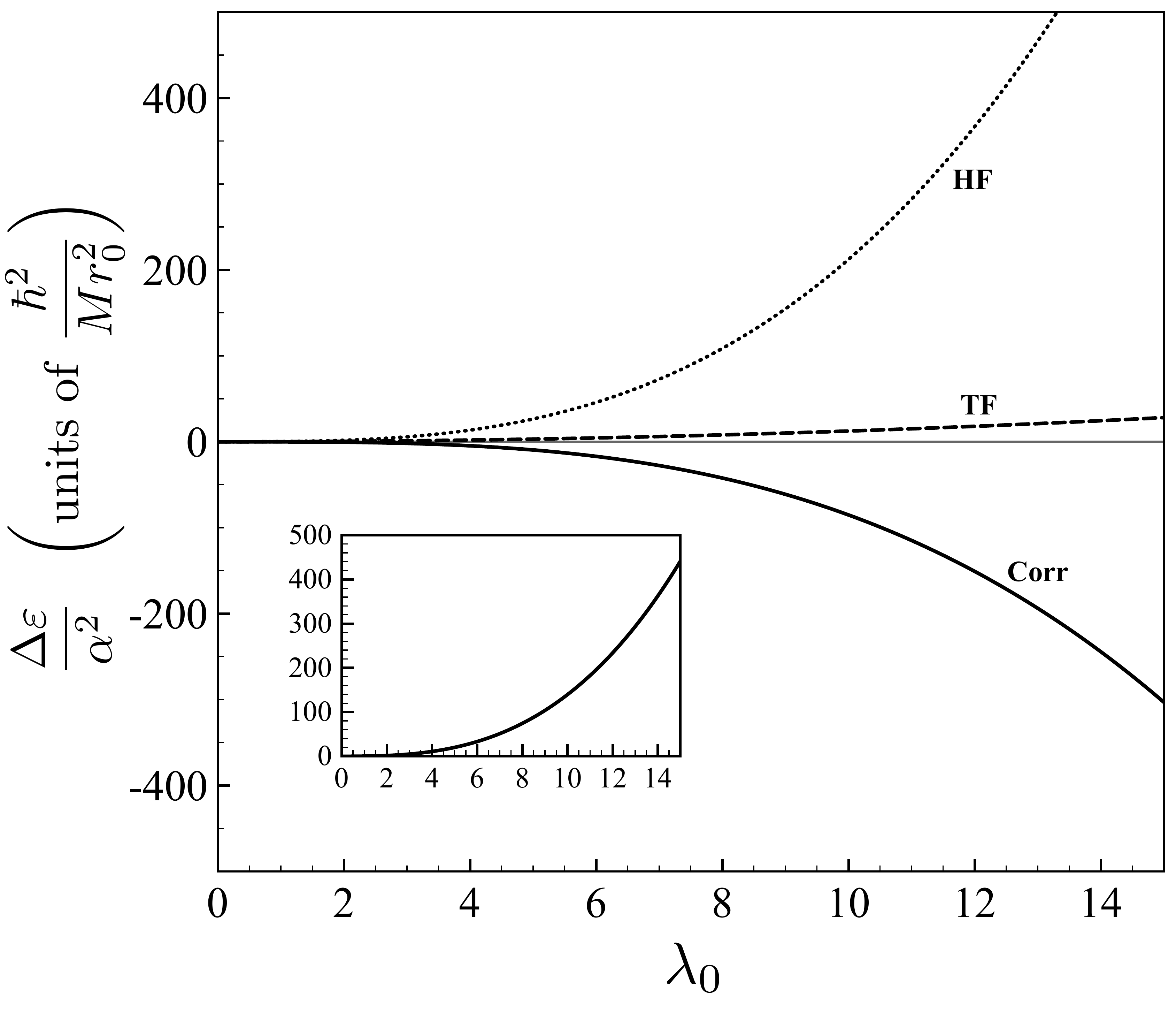}}
\caption{A plot of the three terms in Eq.~\eqref{deltaE2}.  The dashed, dotted and solid curves correspond to the TF, HF and correlation terms,
respectively.  Inset:  The sum of the three terms in Eq.~\eqref{deltaE2}.  The axes in the inset are as in the main figure.}
\label{fig1}
\end{figure}

In Fig.~\ref{fig1}, we present a plot of three terms occurring in Eq.~\eqref{deltaE2}.  The dashed, dotted, and solid curves correspond to the TF, HF and correlation terms,
respectively.  The inset to Fig.~\ref{fig1} displays the sum of the three terms,  
which clearly 
reveals that $\Delta \varepsilon \geq 0$.  This result implies that, at the level of the DFT defined by Eq.~\eqref{EtotLDA}, the FL phase is always stable toward 
a transition to an inhomogeneous density distribution.
In order to understand why the present DFT fails to predict a phase transition beyond
a critical coupling, 
we need to revisit our construction of the total energy functional for an inhomogeneous
system.  

As an immediate step toward improving the quality of our DFT, we may augment Eq.~\eqref{2dkelambda},
with an {\em ad hoc} von Weizs\"acker-like (vW) gradient correction~\cite{vW,note1}, which explicitly takes into account
the increase in the KE associated with a nonuniform spatial density, {\it viz.,}
\bea\label{2dvW}
K_{\rm vW}[\rho] = \lambda_{\rm vW} \frac{\hbar^2}{8 M} \int d^2r \frac{|\nabla \rho(\br)|^2}{\rho(\br)}~,
\eea
or in terms of $\lambda$ and $r_0$, 
\beq\label{2dvWlambda}
K_{\rm vW}[\lambda] = \frac{\lambda_{\rm vW}}{8\pi} \frac{\hbar^2}{M r_0^2}\int d^2r~ |\nabla \lambda(\br)|^2~.
\eeq
The value of the vW coefficient in 2D typically lies within the range 
$0 < \lambda_{\rm vW} \lesssim 0.05$~\cite{vanZyl12}.  In our numerical calculations, we will take $\lambda_{\rm vW} = 0.0184$, as this is the value it interpolates to
in the thermodynamic limit~\cite{vanZyl12}.  

The vW contribution,
Eq.~\eqref{2dvWlambda}, introduces an additional term to Eqs.~\eqref{deltaE2} and \eqref{deltaE2WC}.  Specifically, Eq.~\eqref{lambdamod} leads to 
\bea\label{deltaEvW}
\frac{\Delta \varepsilon_{\rm vW}}{\alpha^2} &=&
\lambda_{\rm vW} \frac{(r_0q)^2}{16}~,
\eea
whereas Eq.~\eqref{lambdaWC} gives
\bea\label{deltaEvWWC}
\frac{\Delta \varepsilon_{\rm vW}}{\alpha^2} &=&
\lambda_{\rm vW} \frac{3(r_0q)^2}{4}~.
\eea
The dependence
on $q$ in the vW correction is characteristic of going beyond the LDA, {\it i.e.,} $q\neq 0$.  While the inclusion
of the vW functional to the TF KE is known to provide smooth equilibrium density distributions~\cite{vanZyl12}, and a good description of the collective modes of the 2D dFG~\cite{vanZyl14}, 
its resulting {\em positive} contribution to Eqs.~\eqref{deltaE2} and \eqref{deltaE2WC}
does not alter the results gleaned from Figure~\ref{fig1}.  In other words, a gradient correction to the TF KE functional
offers no remedy for the absence of a phase transition. 

Next, we examine more carefully the HF contribution to $E_{\rm int}[\rho]$ in the case where
the 2D dFG is inhomogeneous.  As mentioned above, it is generally accepted in most situations that the HF energy, Eq.~\eqref{E1dd},  for the uniform system, may be used
within the LDA for developing a DFT for  investigating inhomogeneous systems.  
However, in the present case, 
Eq.~\eqref{E1dd} alone is clearly insufficient.  
Indeed, when dealing with an inhomogeneous 2D dFG, the HF energy also has an inherently {\em nonlocal}
contribution, which up to now we have ignored.  

The nonlocal
piece to the HF energy is given by~\cite{fang,vanZyl12}
\bea\label{hfE2}
E_{\rm dd}^{(2)}[\rho] &=&- \frac{C_{\rm dd}}{4}\int d^2r~ \rho(\br) \int d^2r' \int \frac{d^2k}{(2\pi)^2}~ k e^{-i\bk \cdot (\br-\br')}
\rho(\br')\nonumber \\
&=& -\frac{1}{4}C_{\rm dd} \int \frac{d^2k}{(2\pi)^2} k |\bar{\rho}(\bk)|^2~,
\eea
where $\bar{\rho}(\bk)$ is the Fourier transform of $\rho(\br)$.
Note that Eq.~\eqref{hfE2} vanishes in the uniform limit, while its negative sign serves to crucially {\em lower} the total energy of the system when the density is non-uniform.  

Inserting Eq.~\eqref{denmod} into
Eq.~\eqref{hfE2} gives
\bea\label{epsilon2dd}
\frac{\Delta \varepsilon^{(2)}_{\rm dd}}{\alpha^2} &=& -\frac{1}{8} (r_0q) \lambda_0^2~,
\eea
while using Eq.~\eqref{denmodWC} in \eqref{hfE2} yields
\bea\label{epsilon2ddWC}
\frac{\Delta \varepsilon^{(2)}_{\rm dd}}{\alpha^2} &=& -\frac{3}{2} (r_0q) \lambda_0^2~.
\eea
The $q$ dependence in Eqs.~\eqref{epsilon2dd} and \eqref{epsilon2ddWC} is again indicative of the nonlocality of the theory.
Including contributions from the vW  and nonlocal HF
energies  leads to a modified energy difference.
For the density modulation specified by Eq.~\eqref{denmod}, we obtain
\beq\label{deltaE3}
\frac{\Delta \tilde{\varepsilon}}{\alpha^2} = \frac{1}{8}\lambda_0^2 +\lambda_{\rm vW} \frac{\tq^2}{16} +
\left[ \frac{2}{3\pi} - \frac{1}{8} \frac{\tq}{\lambda_0} \right]      \lambda_0^3
 -\frac{1}{8}\lambda_0^4 \left[ \frac{3}{2} \ln [f_0] + \frac{11}{16}\lambda_0 A + \frac{1}{16} \lambda_0^2 B\right]~,
\eeq
where we have defined $\tilde{q}=q r_0$.
Similarly, the triangular symmetric density modulation of Eq.~\eqref{denmodWC} gives
\beq\label{deltaE3WC}
\frac{\Delta \tilde{\varepsilon}}{\alpha^2} = \frac{3}{2}\lambda_0^2 +\lambda_{\rm vW} \frac{3 \tq^2}{4} +
\left[ \frac{8}{\pi} - \frac{3}{2} \frac{\tq}{\lambda_0} \right]      \lambda_0^3
 -\frac{1}{4}\lambda_0^4 \left[ 9 \ln [f_0] + \frac{33}{8}\lambda_0 A + \frac{3}{8} \lambda_0^2 B\right]~.
\eeq
We will now use Eqs.~\eqref{deltaE3} and \eqref{deltaE3WC} to study the transition to a 1DSP and triangular WC.


\subsection{1D stripe phase and triangular Wigner crystal}\label{stripes}

\begin{figure}[ht]
\centering \scalebox{0.4}
{\includegraphics{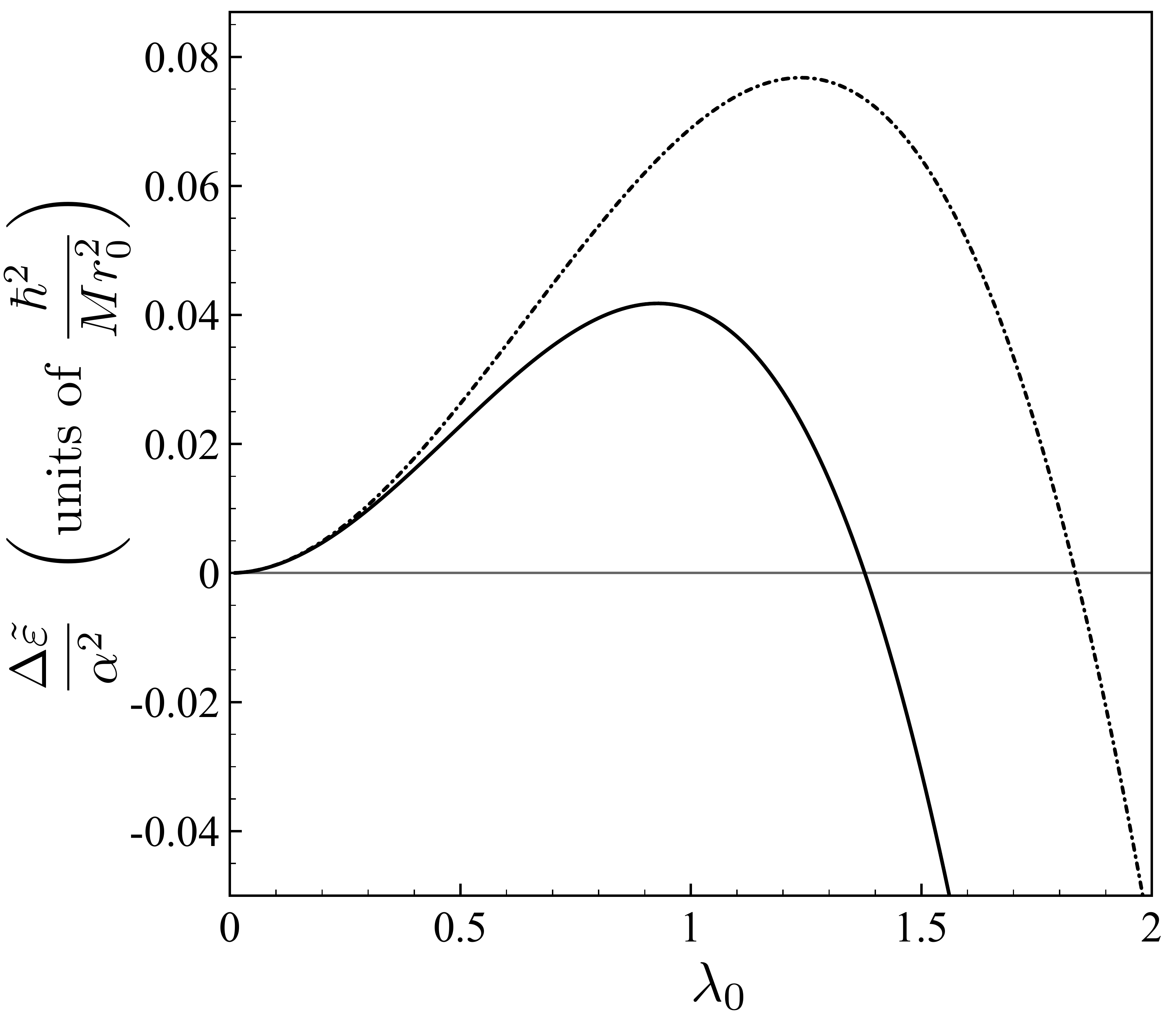}}
\caption{The modified energy difference, Eq.~\eqref{deltaE3}, for the 1D stripe phase (solid curve) and triangular WC (dot-dashed curve).  The transition to a 1D stripe phase occurs at
$\lambda_0 \approx 1.38$ while for the triangular WC,  $\lambda_0 \approx 1.84$. }
\label{fig2}
\end{figure}

We begin by investigating the possible transition to a 1DSP.  Specifically,  we consider
Eq.~\eqref{deltaE3} under 
a density-wave modulation, Eq.~\eqref{denmod}, with wave vector $\tilde{\bq} = 2 k_Fr_0\hat{y}$, since it is expected to have the lowest energy cost for the formation of the 
stripe phase~\cite{Sieberer11,Block12,Babadi11,block2014,note1a}.  
The solid curve in Fig.~\ref{fig2} depicts the energy difference between
the 1D stripe and the uniform phase as $\lambda_0$ is varied.   We note that $\Delta \tilde{\varepsilon}$ crosses zero,
indicating that the stripe phase has lower energy than the uniform phase for $\lambda_0=k_Fr_0 \gtrsim 1.4$.  The
fact that $\Delta \tilde{\varepsilon}$ changes sign also nicely 
emphasizes the importance of including the nonlocal HF energy, $E^{(2)}_{\rm dd}$, to account for an instability in
the liquid phase. 
Our result for the onset of the transition compares well with other theoretical  approaches~\cite{Sieberer11,Block12,Babadi11,block2014}
which all yield a value of $\lambda_0 \approx 1.4$, in agreement with
our DFT prediction.   One may be tempted to believe that the aforementioned agreement is somewhat fortuitous; after all, the vW coefficient, $\lambda_{\rm vW}$, is still an adjustable parameter.
However, even if we set $\lambda_{\rm vW}=0$, we obtain $k_{\rm F}r_0 \approx 1.3$, which is still in good agreement with earlier results.  
Moreover, in the extreme limit of $\lambda_{\rm vW}=1$, (which is well outside of the realm of realistic values, $0 <  \lambda_{\rm vW} \lesssim 0.05$, discussed in Ref.~\cite{vanZyl12}), the transition is only shifted
to $k_{\rm F}r_0 \approx 3.4$.   Regardless, adjusting $\lambda_{\rm vW}$ within the range $0 < \lambda_{\rm vW} \lesssim 0.05$, has no significant impact on the location of the transition.

Following Ref.~\cite{ghosh95}, one may also attempt to use Eq.~\eqref{denmod} to investigate the transition to a triangular WC phase.  In this case, the wave vector, $\tq$, is related to the density
by demanding only one atom per primitive cell in a triangular lattice.  We readily find that
\beq\label{qtr}
\tq = q r_0= \left (\frac{8\pi}{3^{1/2}}\right)^{\frac{1}{2}}\frac{\lambda_0}{2}~.
\eeq
Using Eq.~\eqref{qtr} in Eq.~\eqref{deltaE3} leads to the dot-dashed curve in Fig.~\ref{fig2}, which changes sign at $k_{\rm F}r_0 \approx 1.84$.  We therefore conclude that the 1DSP is always the energetically
 favoured ordered phase, at least within the confines of the density modulation {\em ansatz}, Eq.~\eqref{denmod}.


\subsection{Triangular and square Wigner crystal}\label{wigner}

\begin{figure}[ht]
\centering \scalebox{0.4}
{\includegraphics{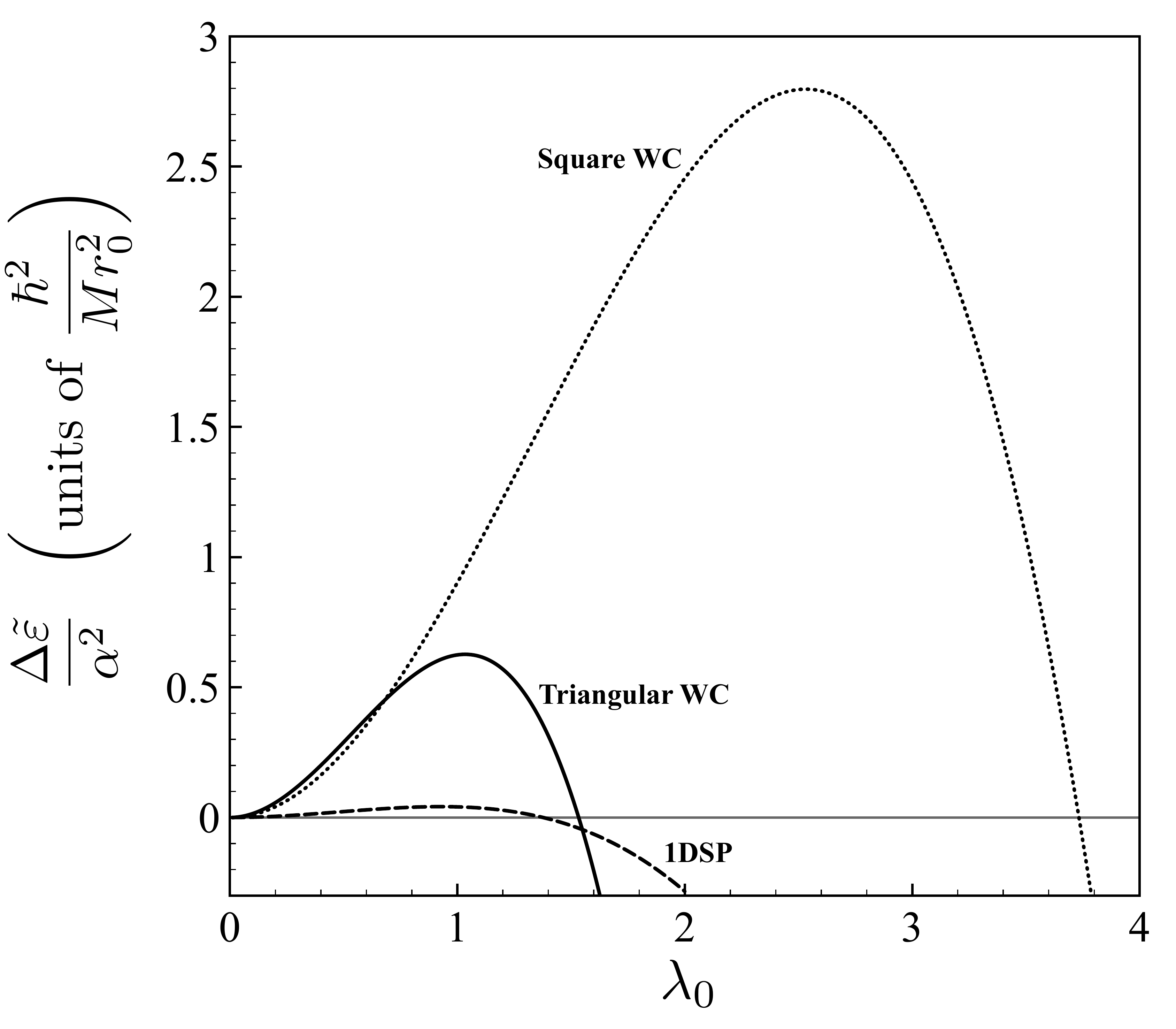}}
\caption{The modified energy difference, Eq.~\eqref{deltaE3WC}, for the triangular WC (solid curve).  The transition to a WC occurs at
$\lambda_0 \approx 1.52$.  The dashed curve is taken from Fig.~\ref{fig2} (solid curve in Fig.~\ref{fig2}). The dotted curve is the modified energy difference for a square WC, Eq.~\eqref{deltaE3SQ}.
 }
\label{fig3}
\end{figure}

The results for the WC phase obtained above can be improved upon by using Eq.~\eqref{deltaE3WC}, which we recall was derived by employing
the more realistic triangular symmetric density modulation, Eq.~\eqref{denmodWC}.  We will use Eq.~\eqref{qtr} for the wave vector~\cite{note1a} 
in Eq.~\eqref{deltaE3WC} with $\lambda_0$ being varied.
The solid curve in Fig.~\ref{fig3} indicates that there is a transition to a triangular WC at $k_{\rm F}r_0\approx 1.52$, which lies slightly below the value using the density modulation, Eq.~\eqref{denmod}.
However, Eq.~\eqref{denmodWC} is a much better representation for the WC phase, so we believe that the value $k_{\rm F}r_0\approx 1.52$ is more trustworthy in the context of our weakly modulated
density profiles.  
The dashed curve in Fig.~\ref{fig3}
is taken from Fig.~\ref{fig2} (where it is represented by the solid curve), and is included to allow us to compare the relative energies of the two ordered phases.  We observe that the 1DSP
transition occurs {\em before} the WC, but as we increase the coupling strength, $\lambda_0$, the WC phase becomes the energetically favourable ground state.  We note that there is
only a very small window in which the 1DSP is the preferred ordered state, suggesting that an experimental verification of our results may be difficult.
It is also important to mention that our location for the WC transition, $k_{\rm F}r_0\approx 1.52$, is significantly lower than predicted in Ref.~\cite{babadi2013} and Refs.~\cite{Mateeva12,abed12}, 
which give values of $\lambda_0 = 29 \pm 4$ and $\lambda_0 = 25\pm 3$, respectively.  Nevertheless, similar to what was found in Ref.~\cite{Mateeva12}, the difference in energy between the
1DSP and WC is quite small in the vicinity of the WC transition.

It is  a useful check of our DFT to briefly investigate the
case of a square lattice.  In particular, we expect the square WC to have a higher energy cost compared to either the 1DSP or the triangular WC.  In order to illustrate that our approach
does indeed correctly capture this notion, we show in Fig.~\ref{fig3} (dotted curve) the results of a calculation for the square WC, {\it viz.,}
\beq\label{square density}
\rho(x,y) = \rho_0\left(\sqrt{1-\alpha^2} + \alpha (\cos(qx)+\cos(qy))\right)^2~,
\eeq
and the associated modified energy difference,
\beq\label{deltaE3SQ}
\frac{\Delta \tilde{\varepsilon}}{\alpha^2} = \lambda_0^2 +\lambda_{\rm vW} \frac{\tq^2}{2} +
\left[ \frac{16}{3\pi} -  \frac{\tq}{\lambda_0} \right]      \lambda_0^3
 -\frac{1}{8}\lambda_0^4 \left[ 12 \ln [f_0] + \frac{11}{2}\lambda_0 A + \frac{1}{2} \lambda_0^2 B\right]~,
\eeq
with $\tq = q r_0= \sqrt{2}\lambda_0$.
It is clear that our DFT does indeed
correctly capture the well-known fact that the square lattice  is higher in energy than either the 1DSP or the triangular lattice configuration.  It is evident from  Fig.~\ref{fig3} that the
square lattice will never be the favoured ordered state given the possibilities of forming either a 1DSP or a triangular WC phase.

It is difficult to pin down exactly why our critical coupling strength for the WC transition is in such disagreement with the QMC and variational approaches.  
Keeping in mind that our density modulations are both smooth, and very weak, we are not resolving the ``high granularity'' of the particle density of the system in the WC phase.  As a result, we are only able to indicate
that a transition to an ordered WC phase is energetically favourable, so the lack of
quantitative agreement with the discrete QMC calculations is perhaps not so surprising. On the other hand, the 1DSP is better 
suited to our smooth density modulation scheme, which may explain the good agreement with previous
calculations.    In the following subsection, we will investigate if choosing a density distribution highly  localized  at each lattice site significantly changes our WC transition.


\subsection{``Granular'' Gaussian density}\label{granular}
In order to establish if a different density profile for the triangular crystalline phase has a significant affect on the transition, we use a Gaussian density {\it ansatz}, and perform a non-perturbative
analysis ({\it i.e.,} there is no small parameter associated with a weak density modulation) for the energy difference between the liquid and crystal phases.  Specifically, we 
consider the Bravais lattice vectors of the triangular lattice,
\beq\label{a1a2}
\ba_1 = a(1,0),~~~\ba_2=a\left(\frac{1}{2},\frac{\sqrt{3}}{2}\right)~,
\eeq
and the associated reciprocal lattice vectors,
\beq\label{k1k2}
\bk_1 = \frac{2\pi}{a}\left(1,-\frac{1}{\sqrt{3}}\right),~~~\bk_2=\frac{2\pi}{a}\left(0,\frac{2}{\sqrt{3}}\right)~.
\eeq
Here, the lattice constant, $a$, is linked to the density by requiring one atom per unit cell, {\it viz.,}
\beq\label{lattconst}
a = \sqrt{\frac{2}{\sqrt{3}\rho_0}}~.
\eeq
In terms of $\lambda_0$ and $r_0$, we have
\beq\label{lattconst}
a = \sqrt{\frac{8\pi}{\sqrt{3}}}\frac{r_0}{\lambda_0}~.
\eeq
The unit cell itself is defined by the region,
\bea\label{unitcell}
y=\sqrt{3}x~,~~y=\sqrt{3}x-\sqrt{3}a~,~~y=0~,~~y=\frac{\sqrt{3}}{2}a~.
\eea
Next, we define our ``granular'' density distribution with triangular symmetry, {\it viz.,}
\beq\label{densum}
\rho(\br) = \frac{\alpha}{\pi} \sum_{m,n}e^{-\alpha|\br - {\bf R}_{mn}|^2}~,
\eeq
where ${\bf R}_{mn} = m \ba_1 +n \ba_2 $. The quantity $\alpha$ is a localization parameter, and  $m$ and $n$ are positive or negative integers, including zero.  Note that we have chosen a simple form, where  each Gaussian
is isotropic, which is well justified close to the positions of the Bravais lattice vectors~\cite{teeffelen08}.  Equation~\eqref{densum} may also be written as a summation of the reciprocal lattice vectors,
{\it viz.,}
\beq\label{densumk}
\rho(\br) = \rho_0 \sum_{m,n} e^{-k_{mn}^2/4\alpha} e^{i \bk_{mn}\cdot\br}~,
\eeq
where $\bk_{mn} = m \bk_1 + n \bk_2$.

We now define the following dimensionless quantities, $\tilde{\alpha} = \alpha/\rho_0$, $\tilde{a}=a\sqrt{\rho_0}$, $\tilde{x}=x\sqrt{\rho_0}$,
and $\tilde{y} = y\sqrt{\rho_0}$.  
Equation~\eqref{densum} then reads,
\beq\label{densumscaled}
\rho(\tilde{\br}) = \frac{\talpha\rho_0}{\pi} \sum_{m,n}e^{-\talpha|(\tx,\ty)-\sqrt{\frac{2}{\sqrt{3}}}[m(1,0)+n(1/2,\sqrt{3}/2)]|^2}~,
\eeq
which can be written as 
\bea\label{dentri3}
4\pi r_0^2 \rho(\tilde{\br}) &=& 4\pi r_0^2 \rho_0 \frac{\tilde{\alpha}}{\pi}\sum_{m,n}e^{-\talpha|(\tx,\ty)-\sqrt{\frac{2}{\sqrt{3}}}[m(1,0)+n(1/2,\sqrt{3}/2)]|^2}~,
\eea
or finally,
\bea\label{lambdatri}
\lambda(\tilde{\br};\talpha) &=& \lambda_0\left[\frac{\tilde{\alpha}}{\pi}\sum_{m,n}e^{-\talpha|(\tx,\ty)-\sqrt{\frac{2}{\sqrt{3}}}[m(1,0)+n(1/2,\sqrt{3}/2)]|^2}\right]^{\frac{1}{2}} ~.
\eea
The region defining the unit cell, now in terms of dimensionless quantities, reads
\bea\label{unitcell2}
\ty=\sqrt{3}\tx~,~~\ty=\sqrt{3}\tx-\sqrt{2\sqrt{3}}~,~~\ty=0~,~~\ty=\sqrt{\frac{\sqrt{3}}{2}}~.
\eea
The total energy for the inhomogeneous system is then given by Eq.~\eqref{EtotLDA}, supplemented with Eq.~\eqref{hfE2}.   It is useful to note that Eq.~\eqref{hfE2} can be solved analytically by using Eq.~\eqref{densumk}.  The final result of such a calculation
leads to the  nonlocal piece of the HF energy,
\beq
\varepsilon^{(2)}_{\rm dd} = -\left(\frac{\pi}{8\sqrt{3}}\right)^{\frac{1}{2}} \lambda_0^3\sum_{m,n} \sqrt{m^2 -mn+n^2}~e^{-\frac{4\pi^2}{\sqrt{3}\talpha}(m^2-mn+n^2)}~.
\eeq
\begin{figure}[ht]
\centering \scalebox{0.4}
{\includegraphics{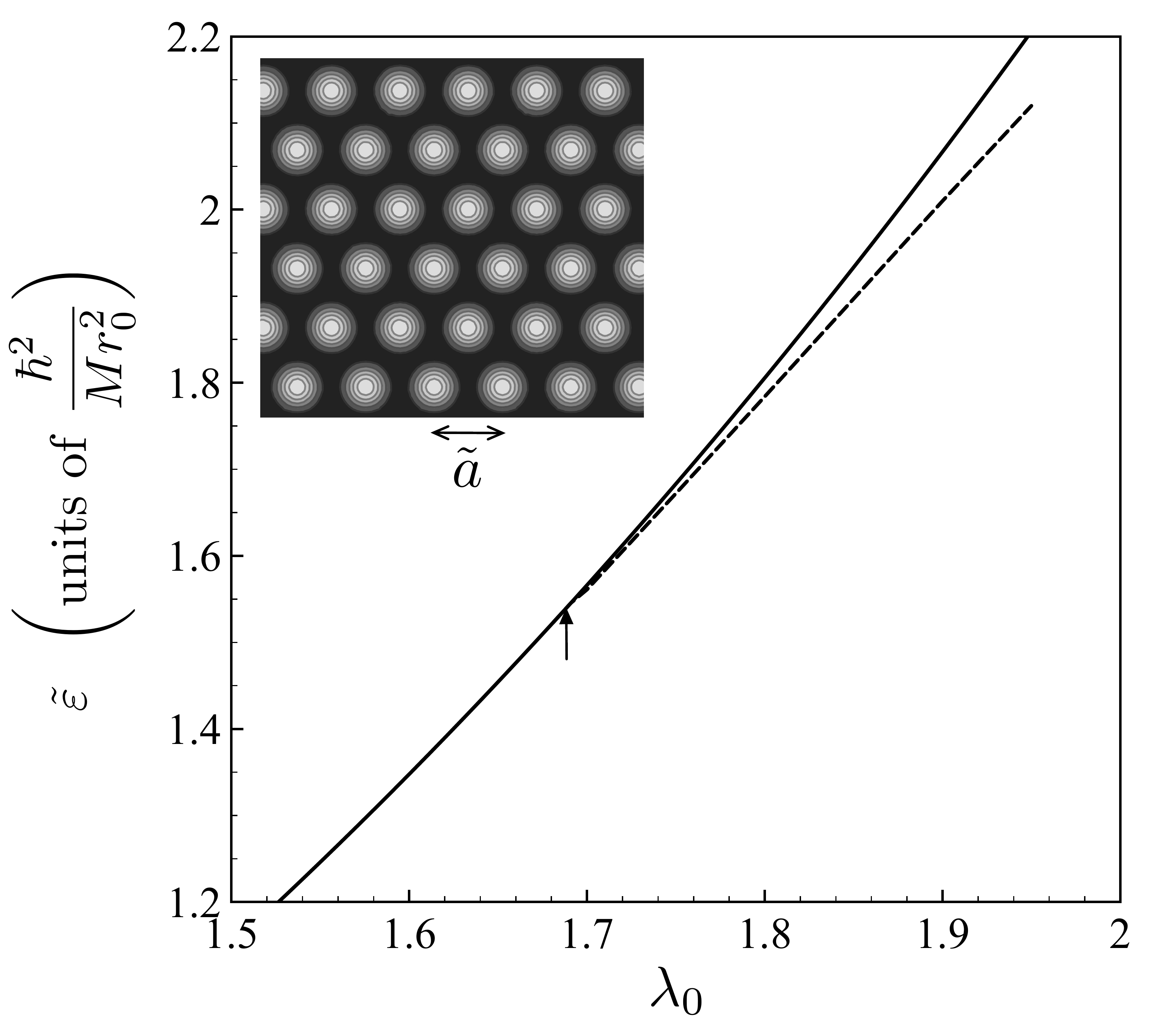}}
\caption{The energy per particle for the uniform phase (solid curve) and the triangular WC (dashed curve).  The transition to a WC occurs at
$\lambda_0 \approx 1.68$, indicated by the vertical arrow in the figure.   Inset: A contour plot of the density distribution for $\lambda_0=1.8$ and $\talpha=8.3$ (corresponding
to the minimum in the total energy).  White: maximal density.  Black: vanishing density.  The dimensionless
lattice constant is $\tilde{a} = \sqrt{\frac{2}{\sqrt{3}}}$.}
\label{fig4}
\end{figure}

 We  numerically investigate the
total energy per particle,
\beq\label{Etot}
\tilde{\varepsilon}(\talpha) = \frac{E[\lambda(\tilde{\br};\talpha)]}{\int d^2r \rho_0}~,  
\eeq
of the ordered phase as follows.
For a {\em fixed} $\lambda_0$, we calculate the total energy, with $\talpha$ as a variational parameter.  We look for a minimum in ${\tilde  \varepsilon}(\talpha)$
for some $\talpha$, and use that as the energy for the inhomogeneous phase.  If the minimum is at $\talpha=0$, the system is in the liquid state.

The findings from this numerical investigation are summarized in Fig.~\ref{fig4}, where we have taken $\lambda_{\rm vW}= 0.0184$~\cite{note2}.  
The solid and dashed curves are the total energy per particle of the uniform and inhomogeneous phases, respectively.
We note that at $\lambda_0=k_{\rm F}r_0 \approx 1.68$, there is a bifurcation, indicating that a transition to the WC takes place ({\it i.e.,} the energy of the triangular lattice is lower than the liquid
phase). We can compare this transition location to $\lambda_0\approx 1.84$ and $\lambda_0 \approx 1.52$ found in Sec.~\ref{stripes} and Sec.~\ref{wigner}, respectively.  We do not believe that
there is any significance to the fact that  $\lambda_0=1.68$ lies exactly in the middle of the previous transition locations.  We conclude that
the precise form of the density profile does affect the location of the transition, although it does not alter the fact that our DFT predicts the formation of a 1DSP {\em before} the formation of a WC.  It is also evident that in spite of using a localized density distribution at each lattice site (see inset to Fig.~\ref{fig4}), our location for the transition to a WC is still in 
drastic disagreement with Refs.~\cite{babadi2013,Mateeva12,abed12}.  

To gain some additional insight into this discrepancy, it is instructive to consider the limiting case of ``point'' dipoles arranged on a triangular lattice ({\it i.e.,} this would correspond to the large $\alpha$ limit in Eq.~\eqref{densumk}).  We can then calculate
the total potential energy per dipole, and compare it to what is obtained in DFT in the same limit.  For the triangular lattice, the total potential energy per ideal dipole is (units of $\hbar^2/Mr_0^2$)
\beq\label{Udd}
U_{\rm dd} = \frac{1}{2}\left( \frac{\sqrt{3}}{8\pi}\right)^{\frac{3}{2}} \lambda_0^3 \sum_{m,n}{}^{'}Å\frac{1}{(m^2 + mn + n^2)^\frac{3}{2}}\approx 5.52~\left( \frac{\sqrt{3}}{8\pi}\right)^{\frac{3}{2}} \lambda_0^3~, 
\eeq
where the primed summation denotes omission of the $m=n=0$ term.  We note that $U_{\rm dd}>0$, as expected for repulsive dipolar interactions.  
However, in our DFT, the HF energy, $E^{(1)}_{\rm dd}+E^{(2)}_{\rm dd}$, dominates in the high density, localized limit, and leads to an unphysical
divergence of the interaction energy to negative values.  For this reason, we cannot go beyond $\lambda_0=2$ for the inhomogeneous system in Fig.~\ref{fig4}, since a minimum 
in $\tilde{\varepsilon}(\talpha)$ is no longer found for any $\talpha$; the implication being that the system is unstable to the formation of a WC for {\em any} $\talpha \neq 0$.
The diverging negative energy likely arises from the fact that the LDA to the  HF energy, $E^{(1)}_{\rm dd}>0$, is being severely underestimated in the localized limit.  
On the other hand, 
$E^{(2)}_{\rm dd}<0$ has no approximations in its form, and is not subject to the LDA.
We therefore suggest that the large discrepancy between the QMC and DFT predictions for the location of the WC transition may be in part attributed to the break-down of the LDA for the
HF energy functional in the highly localized limit.  In fact, there is a delicate balance between the positive and negative
energy contributions to the total energy, and a relatively small change to one of the functionals can cause a large
shift for the critical $\lambda_0$ of the WC transition.

\section{Conclusions and Closing Remarks}\label{closing}

We have presented a DFT for a 2D dFG,  and applied it to examine the instability of the normal FL to an ordered phase.
In Secs.~\ref{stripes} and~\ref{wigner}, a perturbative approach was used, and it was  found that a 1DSP forms at $k_{\rm F}r_0 \approx 1.38$, followed by a transition to a triangular WC
at $k_{\rm F}r_0 \approx 1.52$.  While our prediction
for the onset of the 1DSP is in agreement with other theoretical calculations, our value for the onset of the WC is an order of magnitude smaller than estimates based on variational and QMC
calculations.  In Sec.~\ref{granular}, a highly localized density distribution at each site of the triangular lattice was used to investigate if the transition to a WC could be brought into better agreement
with the QMC results.  Unfortunately, even with this more realistic density profile, the order of magnitude discrepancy for the location of the WC between our DFT and QMC calculations cannot
be resolved.  We suggest that further tests of the efficacy of the LDA used for the HF energy functional need to be performed to determine if it is the root cause of the large discrepancy.
Regardless, we are confident that our DFT result, indicating that a 1DSP precedes the formation of a triangular WC, is qualitatively correct given that the perturbative calculations 
in Secs.~\ref{stripes} and~\ref{wigner} do not probe the
highly localized limit, where the LDA may be in peril.

One of the other significant aspects of this work was showing that the nonlocal part of the HF energy is absolutely crucial for the onset of the density instability.  This is an important point, given that in  
some energy functional based approaches (see {\it e.g.,} Refs.~\cite{Yamaguchi10,bruun11}), the nonlocal HF energy is completely ignored; that is, the total energy functional of the {\em uniform} system
(which manifestly ignores the nonlocal HF term) is used for investigating inhomogeneous systems.  As a result,  instabilities only arise from the anisotropic 
dipolar interaction, which can become attractive when the moments are canted at an appropriate angle relative to the $z$-axis.  Along these lines, it would be of great interest to extend the present DFT to be able to
deal with a fully anisotropic 2D dipolar interaction, and construct the phase diagram of the instabilities  in both the repulsive and attractive regimes.  In addition, including an
external potential is, in principle, straightforward in DFT, thereby opening up the possibility of studying the influence of magneto-optical traps on the density instabilities studied in this paper.
Finally, we plan on extending the present work to include an examination of the affect of temperature on the formation of the stripe and Wigner crystal phases.

\acknowledgements
This work was supported by grants from the Natural Sciences and Engineering Research Council of Canada (NSERC). W. Kirkby and W. Ferguson would like to thank the NSERC Undergraduate Summer Research
Award (USRA) for additional financial  support.  BvZ would like to thank E. Taylor for suggesting this problem, and for useful discussions during the early stages of this work.


\begin{thebibliography}{99}
\bibitem{note1a} One may ask why $\tq$ is not being used as a variational parameter in order to determine the minimum of the total energy.  Unfortunately, such a variational
calculation is only meaningful if the vW gradient correction is present (see Eqs.~\eqref{deltaE3} and~\eqref{deltaE3WC}).  Owing to the fact that the vW correction {\em is not}
essential to our approach, we have chosen
to view $\tq$ as an input parameter, from which the transition can be determined.  More importantly, the $\tq$ leading to a minimum in the energy is highly sensitive to the value of
$\lambda_{\rm vW}$, {\it i.e.,} $\tq_{\rm min} = \lambda_0^2/\lambda_{\rm vW}$, which is not a desirable feature in a variational calculation.
\bibitem{note2} Using $\lambda_{\rm vW}=1$ only shifts the transition up to $\lambda_0 \approx 4$, which is still an order of magnitude smaller than the QMC results.
\end{thebibliography}
\end{document}